\documentclass[12pt]{article}
\usepackage{epsf}
\usepackage{a4}
\newcommand{\ee}{\end{equation}}
\newcommand{\be}{\begin{equation}}
\newcommand{\ea}{\end{eqnarray}}
\newcommand{\ba}{\begin{eqnarray}}

\begin{document}
\thispagestyle{empty}
\begin{flushright}
LU TP 98-3\\
NORDITA 98/5 N/P\\
hep-ph/9801418\\
January 1998\\
\end{flushright}

\vspace{2cm}
\begin{center}
\begin{Large}
Chiral Corrections to Vector Meson Decay Constants
 \\[1cm]
\end{Large}
J. Bijnens$^1$, P. Gosdzinsky$^{2}$ and P. Talavera$^3$ \\[2cm]
${}^1$ Dept. of Theor. Phys., Univ. Lund, S\"olvegatan 14A, S--22362 Lund.\\
${}^2$ NORDITA, Blegdamsvej 17, DK-2100 Copenhagen \O , Denmark. \\
${}^3$ Dept. de F{\'\i}sica i Enginyeria Nuclear,
UPC, E-$08034$ Barcelona, Spain.\\[1cm]
January 1998\\[1cm]
{\bf Pacs:} 11.30Rd, 12.39.Fe, 13.40.Hq, 14.40.Cs, 13.35.Dx\\[0.2mm]
{\bf Keywords:} \begin{minipage}[t]{9.5cm} Chiral Symmetry,
Chiral Perturbation Theory,\\Heavy Vector Meson Theory.\end{minipage}
\end{center}
\begin{abstract}
We calculate the leading quark mass corrections of order $m_q\log(m_q)$,
$m_q$ and $m_q^{3/2}$ to the vector meson decay constants within
Heavy Vector Meson Chiral Perturbation Theory. We discuss the issue
of electromagnetic gauge invariance and the heavy mass expansion.
Reasonably good fits to the observed decay constants are obtained.
\end{abstract}
\setcounter{page}{0}

\clearpage

\section{Introduction}
In this letter our aim is to determine within the Heavy Meson Effective
Theory (HMET) the decay constants for the vector nonet. HMET was
introduced \cite{PRL} as the non--relativistic limit
of an interacting theory between vector mesons (heavy mesons) and a
pseudoscalar meson background (light mesons). The theory is formulated
in terms of operators involving the hadronic fields. The main
reason to introduce such a formalism is to recover a well defined power
counting in small masses and momenta.
This is similar to the Heavy Baryon Chiral Perturbation Theory
\cite{heavybaryon}.

We list here the extra terms needed up to order $p^3$ for the
vector decay constants. All the other relevant terms were already
classified in \cite{NPB} and their coefficients estimated there.
In principle these coefficients or Low-Energy-Constants (LEC's)
should  be determined directly from QCD. We simply fit their
values to experiment, using Zweig's rule to limit the number of relevant
constants, just as was done for the vector meson masses in \cite{NPB}.
Here we fit 5 new data with 3 new parameters.
The poor experimental knowledge in the vector meson sector together with the
rapid increase in the number of LECs when we introduce higher orders in the
effective lagrangian is the main lack of this method.
Some other recent papers using HMET for Vector mesons are~\cite{WD,Rey}.

The vector decay constants are experimentally determined through the
branching ratios of $\rho^0,\omega,\phi\to e^+e^-,\mu^+\mu^-$, via
an electromagnetic current in the matrix element,
or the branching ratios of $\tau^-\to \nu_\tau \rho^-,\nu_\tau K^{*-}$,
via the vector part of the weak
current.

Corrections to the decay constants appear at
order $m_q\log(m_q)$. We calculate here up to order $m_q^{3/2}$.
Two-loop contributions start appearing at order $m_q^2$.

We first discuss our notation and a few definitions. In Sect. \ref{gaugeinv}
we discuss how gauge invariance can be used to connect terms at different
orders in the heavy mass expansion. Using this we then proceed in Section
\ref{terms} to list all the terms in the Lagrangian that are needed.
Then we give our main results, the vector decay constants up to order $p^3$
in the HMET expansion.
We then present the numerical results and compare with the experimental values.

In the appendices we describe the slight extension
needed for the weak currents and  we quote
approximate expressions for the vector isospin states.

\section{Definitions and Notation}

We define here our notation and the basis of chiral
transformations. For a general introduction to Chiral Perturbation Theory
see \cite{CHPTreview}.

Under $SU(3)_L \times SU(3)_R$ global chiral rotations the goldstone
fields can be collected in an unitary matrix field, $U(\phi)$
transforming as:
\be
U(\phi) \rightarrow g_R U(\phi) g_L^\dagger, \qquad
g_L \times g_R \in SU(3)_L \times SU(3)_R
\ee
For the chiral coset space $SU(3)_L \times SU(3)_R / SU(3)_V$ our
choice of coordinates allows us to write:
\be
U(\phi) = \exp {i\frac{\lambda^a \phi_a}{ F}} = u^2(\phi), \quad
\frac{\lambda^a \phi_a}{\sqrt{2}} = \left[
\begin{array}{ccc}
{\pi^0\over\sqrt{2}} + {\eta\over \sqrt{6}} & \pi^+ & K^+ \\ \pi^- & -{\pi^0
\over \sqrt{2}} + {\eta \over \sqrt{6}} & K^0 \\ K^- & \overline K^0 & - {2\eta
\over \sqrt{6}}
\end{array} \right]
\ee
where $F \sim F_\pi = 92.4$ MeV and $u(\phi)$ transforms as
\be
u(\phi) \rightarrow g_R u(\phi) h(\phi,g)^{-1}= h(\phi,g) u(\phi) g_L^{-1},
\ee
with $h(\phi,g)$ the so--called compensator field,
an element of the conserved subgroup $SU(3)_V$.\\
In the nonet case the annihilation modes for
the effective vector meson fields are
collected in a  $3 \times 3$ matrix given by
\be
W_\mu = \lambda_a \varphi^a_\mu = \left[
\begin{tabular}{ccc}
$\displaystyle{ { \omega_\mu + \rho^0_\mu  \over \sqrt{2} }}$, &
$ \rho_\mu^+ $, & $ K_\mu^{*+} $ \\
$ \rho_\mu^- $, &
$\displaystyle{ {\omega_\mu -\rho^0_\mu \over \sqrt{2} }}$, &
$ K_\mu^{*0} $ \\
$ K_\mu^{*-} $, & $ \overline K_\mu^{*0} $, & $ \phi_\mu $
\end{tabular} \right]
\label{defW}
\ee
while its hermitian conjugate, $W_\mu^\dagger $, parametrizes
the creation modes.
In what follows we are only concerned with the transverse components of
the $W_\mu$ fields, i.e. $v\cdot W = v\cdot W^\dagger =0$, $v$ is the
chosen reference velocity for the heavy Vector meson.
The 'longitudinal' component of $W_\mu$ can easily be expressed in terms of
the transverse components. This is due to the fact that a massive
vector meson has 4 components, but only 3 degrees of freedom. This
should be understood in the remainder.

Under chiral symmetry the effective vector fields transform as
\be
W_\mu \rightarrow h(\phi,g) W_\mu h^\dagger(\phi,g)
\qquad
W_\mu^\dagger  \rightarrow h(\phi,g) W_\mu^\dagger
h^\dagger(\phi,g).
\ee

As has been mentioned already, our purpose is to compute the vector
decay constants in the effective theory. They are
defined through the matrix
elements:
\be
\label{defdecay}
\langle 0 \mid \overline q_i \gamma_\mu q_j \mid W \rangle
= F_W \varepsilon_\mu
\ee
for a vector meson state normalized to 1 and
with momentum $m_W v_\mu$. $m_W$ is
the mass of the relevant meson and $\varepsilon_\mu$ its polarization vector.

In order to introduce photons (we extend the formalism to
weak currents in the App. \ref{appendixcurrents}) we use the external
field formalism \cite{CHPTreview}, coupling the pseudoscalar fields
to external hermitian matrix fields, $a_\mu$, $v_\mu$ defined as:
\ba
\label{charges}
&& r_\mu = v_\mu+a_\mu = e Q A_\mu^{ext}+ \ldots  \nonumber\\&&
l_\mu = v_\mu-a_\mu = e Q A_\mu^{ext}+ \ldots
\ea
where Q is the diagonal quark charge matrix,
$Q = (1/3)\mbox{diag}(2,-1,-1)$.

This inclusion of $a_\mu$, $v_\mu$ fields promotes the global
chiral symmetry to a local one, allowing thus to define a covariant
derivative and a connection:
\be
D_\mu U = \partial_\mu U -i r_\mu U +i U l_\mu, \quad
\Gamma_\mu = {1\over 2} ( u^\dagger [\partial_\mu -i r_\mu] u +
u [\partial_\mu -i l_\mu] u^\dagger ).
\label{def2}
\ee
Instead of using the $r_\mu$ and $l_\mu$ fields, we
will use the combination
\be
\label{chinv}
Q_\pm = e \left(u^\dagger Q u \pm u Q u^\dagger\right),
\ee
where $Q_\pm$ transforms as $Q_\pm\to h(\phi,g)Q_\pm h(\phi,g)^\dagger$
under chiral transformations.

Notice that the insertion of external field (photon or a weak current)
does not modify the power counting, i.e ${\cal{O}}(e) = 1$.

\section{Gauge and reparametrization invariance}
\label{gaugeinv}

In this section we discuss briefly the constraints from reparametrization
invari\-ance\cite{Manohar} and the additional constraints from
gauge invariance on the HMET lagrangian.
We will discuss it simply in terms of a single neutral vector meson and
the photon.
It is convenient to start from the relation between the relativistic
and the effective fields \cite{Georgi},
see also the discussion of \cite{NPB}:
\be
V^\mu = {1\over \sqrt{2m_V} } \bigg( e^{-im_Vv \cdot x} W_v^\mu +
e^{im_Vv \cdot x} W_v^{ \dagger \mu } \bigg) + W^\mu_\parallel
\label{delALL}
\ee
where the subindex $v$ denotes the velocity of the heavy meson particle
referred to an inertial observer, $W_v$ has momenta small compared
to $m_V v$ and satisfies $v\cdot W_v=0$.
The longitudinal
component is suppressed by $1/m_V$, see Sect. 4 in \cite{NPB}.
Instead, choosing a second reference
frame related with the first by a Lorentz transformation we
should get the same description of the physics. This fact relates
the expression of the vector field in both frames:
\be
v \rightarrow w = v + q, \qquad
W_v^\nu = e^{-i m_V q \cdot x} \bigg(W_w^\nu + w^\nu {q \cdot W_w}
\bigg)\,.
\label{repa1}
\ee
$q$ is infinitesimal and satisfies $v\cdot q=0$ since $v^2=w^2=1$.
This
determines some ${1}/{m_V}$ coefficients in the chiral expansion,
that can not be modified by non--perturbative corrections\cite{Manohar}.

For the photon we need also to consider gauge
invariance. The lagrangian has to be invariant under
\be
A_\mu (x) \rightarrow A_\mu (x) +\partial_\mu \epsilon (x)
\label{gauge1}.
\ee
where $\epsilon (x)$ is the gauge field. This transformation is in a particular
frame, for fixed $v$.
Relevant momenta are around $\pm m_V v$ and $0$,
we therefore perform a Fourier
decomposition into low
and high momenta in all the fields entering in the gauge transformation:
\ba
A_\mu (x) &=&  e^{-im_Vv\cdot x} \hat A_\mu (x)+
e^{im_Vv\cdot x} \hat A_\mu^\dagger (x)  + \tilde A_\mu (x)
\nonumber \\
\epsilon (x) &=&
 e^{-im_Vv\cdot x} \hat \epsilon (x)+
e^{im_Vv\cdot x} \hat \epsilon^\dagger (x) y + \tilde \epsilon (x)
\label{photDES}
\ea
where $\hat A_\mu (x)$, $\tilde A_\mu (x)$,
$\tilde \epsilon (x)$ and $\hat \epsilon (x)$ are all low momentum
fields. As was mentioned in \cite{match} a similar decomposition
allows to take into account properly the low momentum component
for the vector meson field.

Since the (low momentum) electromagnetic $U(1)$ is a
subgroup of (the low momentum) $SU(3)_V$, the effect of
$\tilde \epsilon (x)$ and $\tilde A_\mu$
will be ignored in what follows, it is treated by
the covariant derivatives defined above. Contrary
the high momentum of the
electromagnetic field, $U(1)$, is not included in that group. In order
to obtain a gauge invariant lagrangian under this high momentum subgroup
the electromagnetic field should transform as
\be
\hat A_\mu (x) \rightarrow \hat A_\mu (x) -i m_V v_\mu
\hat \epsilon (x) + \partial_\mu \hat \epsilon (x).
\label{highgauge}
\ee
The transformation in Eq. (\ref{highgauge}) is the equivalent of
Eq. (\ref{repa1}) for the gauge invariance. As Eq. (\ref{repa1}) it
determines
higher order coefficients in
the $1/m_V$ expansion.

For instance, if we take the following toy lagrangian:
\ba
{\cal L} &=& W_\mu^\dagger
\bigg( \hat A^\mu +\alpha_1 \partial^\mu (v \cdot \hat A) +
\alpha_2 (v\cdot \partial) \hat A^\mu \bigg) + h.c. \nonumber \\
&=& W_\mu^\dagger v_\nu
\bigg( v^\nu \hat A^\mu +\alpha_1 \partial^\mu  \hat A^\nu +
\alpha_2 \partial^\nu  \hat A^\mu \bigg) + h.c.
\label{lagtoy}
\ea
a high momentum gauge invariance transformation Eq. (\ref{highgauge}) implies:
\be
\alpha_1 = -{i \over m_V}, \qquad
\alpha_2 = {i \over m_V}.
\ee
those coefficients cannot by modified by non-perturbative corrections.
Repa\-rametrization invariance then requires in addition the (separately
gauge invariant) term
\be
\frac{1}{m_V^2}\partial^\nu W^{\dagger\mu}
\left(\partial^\nu \hat A^\mu - \partial^\mu \hat A^\nu\right) +h.c.
\ee
with the coefficient fixed. Notice that this is precisely the combination
of terms that the relativistic term $\partial^\nu V^\mu
\left(\partial^\nu A^\mu - \partial^\mu A^\nu\right)$ produces,
taking into account $v\cdot W_v=0$.
So terms that are not gauge invariant at first sight can be made gauge
invariant by adding terms of higher order in the heavy mass expansion.
We will work in a fixed gauge, $v\cdot A=0$ to avoid this complication.

\section{The effective lagrangian}
\label{terms}

To construct the relevant terms in the
lagrangian we use both
\be
v\cdot W_v = 0\mbox{ and }
v \cdot \hat A = 0\,, \label{temporal}
\ee
corresponding to the temporal gauge.
Together with Eq. (\ref{chinv}) which involves vertices with
photons and pseudoscalars, we also use the following
building blocks:
\be
\chi_\pm = u^\dagger \chi u^\dagger \pm u \chi^\dagger u, \quad
u_\mu = i u^\dagger \partial_\mu U u^\dagger,
\ee
where $\chi$ contains the
quark mass matrix, $\chi= 2 B_0 \mbox{diag}(m_u, m_d, m_s)$.

We now
construct the most general structure involving
pseudoscalar, vector meson and photon fields which should be
invariant under Lorentz transformations, chiral transformations
charge conjugation, parity  and time reversal.

Making use of
the lowest order equation of motion, $v\cdot D W_\mu=0$, which
eliminates terms removable by field redefinitions,  and all other constraints,
we have the following
non anomalous lagrangian to leading order in $1/N_c$, with $N_c$
the number of colors and $\langle B \rangle =\mbox{tr}(B)$:
\be
{\cal L}_1 = \lambda_1 \langle  W_\mu^\dagger Q_+ \rangle \hat A^\mu
+  \lambda_2
\langle  W_\mu^\dagger \{ \chi_+ , Q_+ \} \rangle \hat A^\mu
+ h.c.
\label{direct}
\ee
And to next order in $1/N_c$:
\be
{\cal L}_2 = \lambda_3  \langle W_\mu^\dagger Q_+ \rangle
\langle \chi_+ \rangle  \hat A^\mu
+  \lambda_4
\langle W_\mu^\dagger \rangle
\langle  \chi_+  Q_+  \rangle \hat A^\mu
+ h.c.
\label{direct1ONC}
\ee
Notice that in the three flavour case, terms involving only a trace over
$Q_\pm$ vanishes, so they never appear.\\
Terms at next-to-leading order in $N_c$ are Zweig rule suppressed and
we will treat the terms in Eq. (\ref{direct1ONC}) as ${\cal O}(p^4)$.

At lowest order the odd intrinsic parity sector of the lagrangian is given by:
\be
{\cal L}_3 = i g\langle \{W_\mu^\dagger , W_\nu \} u_\alpha
\rangle  v_\beta \epsilon^{\mu \nu \alpha \beta } +
 i\lambda_5 \left(\langle W_\mu^\dagger
\{ Q_+, u_\alpha \} \rangle \hat A_\nu v_\beta
\epsilon^{\mu \nu \alpha \beta } + h.c.\right)\,.
\label{anomal}
\ee

In Eq. (\ref{direct}), Eq. (\ref{direct1ONC})  and
Eq. (\ref{anomal}), all the coupling constants
are real numbers\footnote{We
need to use $C$, $P$ and $T$ separately here to prove this.
The HMET is not a relativistic field theory so the CPT theorem is not valid.
$C$ connects $W_\mu$ and $W_\mu^T$. It is only using both the requirement of
a hermitian lagrangian and $T$, which
also connects $W_\mu$ with $W_\mu^\dagger$,
that we can conclude that the $\lambda_i$ are real.}.

In fact, reparametrization invariance requires the presence
of higher order terms proportional to $g$ and $\lambda_5$.
They only contribute to the vector decay constants at order $p^4$.
The connection between a relativistic formulation and the present one
can be done as in \cite{NPB,match} but the external fields need to be split
up as was done for the photon field in Sect. \ref{gaugeinv}.

\section{Calculation of the Vector Decay Constants}

The Vector Meson leptonic
widths are given by
\be
\Gamma( W \rightarrow l^+ l^-) = {8 \pi \alpha_{em}^2 \over 3}
{F_W^2 \over m_W^5} (m_W^2 + 2 m_l^2) \sqrt{ m_W^2 - 4 m_l^2}\
\label{widthREL}
\ee
for a $\vert \overline q_i q_j \vert$ hadronic system which decays
via $W \rightarrow \gamma \rightarrow l^+ l^-$ with a
$eF_W \hat A_\mu^\dagger W^\mu$ effective coupling.
Otherwise if a weak decay current is involved (i.e.
a $\tau$ lepton decay), the width is determined by
\be
\Gamma( \tau \rightarrow W \nu) = { |V_{CKM}|^2 G_F^2 F_W^2 \over
8\pi}
{ (m_\tau^2 - m_W^2)^2 (m_\tau^2 +2 m_W^2) \over m_W m_\tau^3},
\label{weakwidth}
\ee
where $G_F =\sqrt{2}e^2/ (8\sin^2\theta_W M_{\cal W}^2)$
is the Fermi constant and an effective coupling
$F_W \,[e/(2\sqrt{2}\sin\theta_W)V_{CKM}]\, W^\dagger_\mu \hat {\cal W}_+^\mu$
has been used.
$V_{CKM}$ is equal to $V_{ud}$, $V_{us}$ for $\tau^+$ decay to
$\rho^+\bar\nu_\tau$, $K^{*+}\bar\nu_\tau$ respectively. $m_W$ is the mass
of the vector meson, $M_{\cal W} = 80.3$~GeV  is the mass of the weak
vector boson.
The only difference with the definition (\ref{defdecay}) is that we
use the electromagnetic current for the neutral vector bosons.

\begin{figure}
\begin{center}
\leavevmode\epsfxsize=12cm\epsfbox{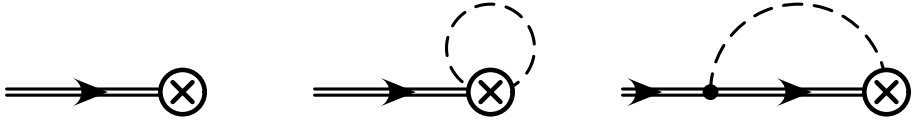}
\end{center}
\caption{\label{diageffec}The three effective diagrams
contributing to the width. A double line is a Vector Meson, a dashed line
a pseudoscalar meson and the circled cross a vertex from ${\cal L}_{1,2,3}$.}
\end{figure}

To determine inside the effective theory
the $F_W$ constants of Eq. (\ref{widthREL}) and (\ref{weakwidth})
one has to compute the diagrams depicted in
Fig. \ref{diageffec} and in addition the contribution coming
from the Vector Meson wave function renormalization (w.f.r.).
For that purpose we first define the physical vector meson basis,
to be used in what follows ($T_{ext}$),
fully diagonalizing the two point Green function ($G_2(p^2,m_{phys}^2$))
at the 1--loop level,
this contribution is found using the results of Ref. \cite{NPB}
for $G_2$ up to ${\cal O}(p^4)$. This
is sufficient to compute the w.f.r. as defined by
\be
Z_V = \frac{\partial}{\partial k_W} G_2^{-1}(k_W,m_{phys})
\mid_{on-shell},
\ee
to order $p^3$, they include the electromagnetic corrections of \cite{BG}.
Here $k_{W\mu} = p_{W\mu}-m_V v_\mu$ and $p_W^2=m_W^2$, $k_W$ is the momentum
in the HMET.

The other contributions are found by direct calculation using
Eq. (\ref{direct}) and Eq. (\ref{anomal}).
We define
\be
\label{defQ0}
\chi_+^0 = \left.\chi_+\right|_{u=0}
 \quad \mbox{ and }\quad Q_+^0 = \left.Q_+\right|_{u=0} \quad
\ee
to be the quantities $Q_+$ and $\chi$ with $u=0$.
The extension of $Q_+$ needed for the charged case is
defined in (\ref{fullQ}).
The contribution of the tree diagram, the first diagram of Fig.
\ref{diageffec},
to the decay constant corresponding to $T^a_{ext}$ is
\footnote{We need to extract a factor of $e$, respectively
$e/(2\sqrt{2}\sin\theta_W)V_{CKM}$, compared to the definition of $Q_+^0$
in (\ref{defQ0}) to obtain the decay constant $F_W$.}
\be
 \lambda_1 \langle T_{ext}^a Q_+^0 \rangle +
 \lambda_2 \langle T_{ext}^a \{ \chi_+^0 , Q_+^0 \} \rangle\,.
\label{treeQED}
\ee
The tadpole type (second) diagram is given by:
\be
\sum_{M =1,8}  {\lambda_1 \over 4 F_\pi^2 }
\langle   T_{ext}^a \big[ T^M , \big[ T^{M \dagger}, Q_+^0
\big] \big] \rangle \mu_M \,,
\label{tadQED}
\ee
and finally the sunrise type diagram (last diagram):
\be
\sum_{M =1,8} \sum_{c=1,9}
{4  g \lambda_5 \over F_\pi^2}
\langle \{ T_{int}^{c\dagger} , T_{ext}^a \} T^M \rangle
\langle T_{int}^c \{ Q_+^0,  T^{ M \dagger} \}  \rangle
K(p_{ext},\Delta m_c, m_M)\,,
\label{jenkinsQED}
\ee
where the $T_{int}$ basis was defined in \cite{NPB} and the $\mu_M$
, $K(p_{ext},\Delta m_c, m_M)$
functions are defined in Eq. (\ref{int}) of App. (\ref{appendixform}).
The full result is then given by the sum of equations (\ref{treeQED}),
(\ref{tadQED}) and (\ref{jenkinsQED}) divided by the square root
of the relevant $Z_V$. We have kept also some higher order terms required by
reparametrization invariance. This corresponds to using instead of
the function $K(p_{ext},\Delta m_c, m_M)$ the full
combination $[G,\Omega]$ defined in (\ref{int2}).

\begin{figure}
\begin{center}
\leavevmode\epsfxsize=4cm\epsfbox{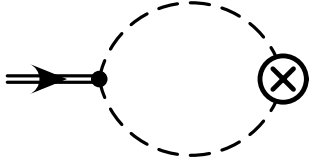}
\end{center}
\caption{\label{diagrel} One of the relativistic diagram
contributing to the decay constant.}
\end{figure}

We have checked  that the non--analytical pieces of the
relativistic diagram in Fig. \ref{diagrel} are fully recovered by
the second diagram in Fig. \ref{diageffec} as was explicitly shown in
\cite{match} for the scalar form--factor case.

\section{Numerical results and conclusions}

{}From the decay widths of Eqs. (\ref{widthREL},\ref{weakwidth})
we obtain the experimental results of the second column of Table
\ref{tablewidths}. The Cabibbo-Kobayashi-Maskawa mixing
angles, lifetimes, masses and branching ratios are taken
from the particle data book \cite{PDG}.
\begin{table}
\begin{center}
\begin{tabular}{|c|c|c|c|c|c|}
\hline
Scenario & exp & LO & III & IV &VI \\
\hline
$100\lambda_1$ (GeV$^{3/2}$) &--&6.42& 6.97(36)  & 7.55(39) & 7.74(38) \\
$100\lambda_2$ (GeV$^{-1/2}$)&--&0.00&$-$0.26(50)& 0.40(49) & 1.21(69) \\
$100\lambda_5$ (GeV$^{1/2}$) &--&0.00& 3.56(75)  & 5.12(71) & 7.24(1.43)\\
\hline
$F_{\rho^0}$ (GeV$^{3/2}$) & 0.0946(23) & 0.0908 & 0.0951 & 0.0955 & 0.0937 \\
$F_{\omega}$ (GeV$^{3/2}$) & 0.0287(5)  & 0.0303 & 0.0286 & 0.0278 & 0.0282 \\
$-F_{\phi}$   (GeV$^{3/2}$) & 0.0565(10) & 0.0428 & 0.0563 & 0.0544 & 0.0534 \\
$F_{\rho^+}$ (GeV$^{3/2}$) & 0.1284(10) & input  & 0.1275 & 0.1234 & 0.1280 \\
$F_{K^{*+}}$ (GeV$^{3/2}$) & 0.1447(45) & 0.1284 & 0.1460 & 0.1571 & 0.1554 \\
\hline
\end{tabular}
\end{center}
\caption{The experimental Decay Constants and various
good fits.
The column labeled LO (lowest order) is
Eq. 
(\protect{\ref{naive}})
with $F_{\rho^+}$ as input.
For an explanation of the
scenarios and the values of the
other constants see Table 1 in
\protect{\cite{NPB}}.
\label{tablewidths}}
\end{table}
The naive prediction, using all the pure isospin states, is
\be
\label{naive}
\sqrt{2}F_{\rho^0}=3\sqrt{2}F_\omega
= -3F_\phi=F_{\rho^+}=F_{K^{*+}}\,.
\ee
This is satisfied to about 5\% for $\rho^+,\rho^0,\omega$, to
about 13\% for $\rho^+,K^{*+}$ and about 32\% for $\rho^+,\phi$.
The signs we have fixed to agree with (\ref{naive}).

The input parameters used are scenarios III,IV and VI from Ref. \cite{NPB}.
III and VI were fits to the masses only for two different values of $g$,
one high and one low, fit IV also included the $\rho-\omega$ mixing in the
input but otherwise as fit III.
We find a reasonable fit to all the decay constants for reasonable values
of $\lambda_{1,2,5}$. The higher order corrections are also
reasonable, below 40\%.
The fit for scenario IV is worse for the following reason:
Using the mixings from \cite{NPB} for the $\omega$, including $p^4$ effects,
the contributions at tree level from the $\lambda_1$ term
essentially cancel, leaving the loop diagrams of Fig. \ref{diageffec}
and $\lambda_2$ as the main contributions. This makes the predictions for
the $\omega$ somewhat unstable. We also have a rather large w.f.r.
factor for the $\phi$. The total size of the higher order corrections can be
judged by comparing the results from Eq. (\ref{naive}), column LO, with
those of the three scenarios.

What we have minimized in order to get the values of $\lambda_{1,2,5}$
in Table \ref{tablewidths} is
$F_{fit}=\sum_{W=1,5} \left(|F^{theo}_W/F^{exp}_W|-1\right)^2$.
The error on the $\lambda_i$ corresponds to  changes in $F_{fit}$ by about
0.01, i.e. at most 10\% for an individual $F_W$, minimizing the other two
$\lambda_i$ at the same time.

In conclusion we have calculated the corrections to the vector decay constants
in heavy vector meson chiral perturbation theory and found acceptable
fits to all the measured ones. We have determined 5 observables in terms
of 3 parameters.

\appendix
\setcounter{equation}{0}
\newcounter{zahler}
\addtocounter{zahler}{1}
\renewcommand{\thesection}{\Alph{zahler}}
\renewcommand{\theequation}{\Alph{zahler}.\arabic{equation}}
\section{Weak currents}
\label{appendixcurrents}
In this appendix we give the main features to incorporate
charged weak current effects to our formalism.
The neutral weak current is not phenomenologically relevant at present.
To do so one needs to extend the left current defined in
Eq. (\ref{charges}) to:
\be
l_\mu = e Q A_\mu + { e\over \sqrt{2} \sin {\theta_W} }
({\cal W}_\mu^\dagger   T_+ + {\cal W}_\mu T_- )
\label{Weak1}
\ee
where $\sin(\theta_W)$ is Weinberg's angle,
${\cal W}_\mu$ parametrizes the spin--1 gauge boson fields
-- it creates an $W^+$ gauge boson field and destroys an $W^-$ one -
and we have introduced the $T$ matrices in terms of the relevant Cabibbo--
Kobayashi--Maskawa factors
\be
T_+ = \left(
\begin{array}{ccc}
0 & V_{ud} & V_{us} \\
0 & 0 & 0 \\
0 & 0 & 0
\end{array}
\right),
\qquad \qquad
T_- = T_+^\dagger  = \left(
\begin{array}{ccc}
0 & 0 & 0 \\
V_{ud} & 0 & 0 \\
V_{us} & 0 & 0
\end{array}
\right).
\ee
As in the QED sector, the requirement of not breaking
chiral symmetry force us to split the ${\cal W}_\mu$ field in different
components in the momenta space according to
\be
{\cal W}_\mu =  e^{-im_Vv\cdot x} \hat {\cal W}_{+ \mu} +
e^{im_Vv\cdot x} \hat {\cal W}_{- \mu}^\dagger,\qquad
{\cal W}_\mu^\dagger =  e^{-im_Vv\cdot
x} \hat {\cal W}_{- \mu} +
e^{im_Vv\cdot x} \hat {\cal W}_{+ \mu}^\dagger\, .
\ee
The inclusion of the charged currents is now achieved by replacing
\be
\label{fullQ}
e Q_+ \hat A_ \mu
\qquad\mbox{ by }\qquad
 e Q_+\hat A_\mu +
{e \over \sqrt{2} \sin  \theta_W }\,
 u (T_+ \hat {\cal W}_{- \mu}^\dagger + T_- \hat {\cal W}_{+ \mu}) u^\dagger
\ee
in lagrangians Eq. (\ref{direct}), Eq. (\ref{direct1ONC}) and Eq.
(\ref{anomal}).
Where now $P$ violation is allowed.

\addtocounter{zahler}{1}
\section{Vector Decay Constant Contributions}
\label{appendixform}
\setcounter{equation}{0}

In this appendix we show the formulae for the contribution
coming from the effective diagrams of Fig. \ref{diageffec}. Where
we take the approximation of non--diagonal fields inside $T_{ext}$ and
$T_{int}$, i.e. we use the pure isospin 1 state for $\rho^0$, pure
isospin 0 for $\omega$ and $\phi$ and the $\phi$ as the pure strange
vector state, i.e. we neglect here $\rho^0-\omega-\phi$ mixing.

For the tree level contribution we find
\ba
\label{tree}
F_{\rho^0} &=& \sqrt{2} \lambda_1 + {8 \sqrt{2} \over 3} B_0 \lambda_2
( m_d + 2 m_u ) \nonumber  \\
F_{\omega} &=& { \sqrt{2} \over 3 } \lambda_1 -
{ 8 \sqrt{2} \over 3} B_0 \lambda_2 (m_d - 2 m_u) \nonumber \\
F_{\phi} &=& -{2 \over 3 } \lambda_1 - {16 \over 3} B_0 m_s \lambda_2
\nonumber  \\
F_{\rho^+} &=&
2 \lambda_1 + 8 B_0 (m_u + m_d) \lambda_2  \nonumber \\
F_{K^{*+}} &=&
2 \lambda_1 + 8 B_0 (m_u + m_s) \lambda_2\,.
\ea

For the tadpole type diagrams, we find
\ba
\label{tadpole}
F_{\rho^0} &=& { \lambda_1 \over \sqrt{2} F_\pi^2}
( \mu_{K^+}   + 2 \mu_{ \pi^+ }) \nonumber  \\
F_{\omega} &=& {\lambda_1 \over \sqrt{2} F_\pi^2} \mu_{K^+} \nonumber \\
F_{\phi} &=& -{\lambda_1 \over F_\pi^2} \mu_{K^+} \nonumber  \\
F_{\rho^+} &=&{ \lambda_1 \over  F_\pi^2 }
\bigg( {\mu_{K^+} \over 2} + { \mu_{K^0} \over 2} +
 \mu_{ \pi^+ } + c^2 \mu_{ \pi^0 }
+ s^2  \mu_\eta  \bigg) \nonumber \\
F_{K^{*+}}
&=& { \lambda_1 \over  F_\pi^2}
\bigg(
 \mu_{K^+}+ {\mu_{K^0} \over 2} +{\mu_{\pi^+} \over 2 } +
( \sqrt{3}  c +s  )^2 { \mu_\eta  \over 4} +
(c - \sqrt{3} s )^2 { \mu_{\pi^0}  \over 4}  \bigg).\nonumber\\
\ea

And finally the contribution coming from the sunrise diagram
\ba
\label{sunri}
F_{\rho^0} &=&  \Lambda   \bigg\{
   \bigg( {c s \over 3 \sqrt{6} }  +
   {s^2 \over \sqrt{2} } \bigg)  [\eta, \omega ] +
   \bigg( {c^2 \over 3 \sqrt{2} }  +
   {c s \over 3 \sqrt{6} } \bigg)  [\eta, \rho^0 ] \nonumber\\&& +
   {1\over 3 \sqrt{2} } [K^+, K^{*+} ] +
   {\sqrt{2} \over 3} [K^0, K^{*0} ] \nonumber  \\
&& + \bigg({c^2 \over \sqrt{2}} - {c s \over 3 \sqrt{6} } \bigg)
     [\pi^0, \omega ] +
  \bigg( {s^2 \over 3 \sqrt{2} } -{c s \over 3 \sqrt{6} }
   \bigg) [\pi^0, \rho^0 ] \bigg\} \nonumber \\
F_{\omega} &=&  \Lambda   \bigg\{
   \bigg( {c^2 \over 9 \sqrt{2} }  +
   {c s \over \sqrt{6} } \bigg) [\eta, \omega ] +
   \bigg( {c s \over \sqrt{6} } +
   {s^2 \over 3 \sqrt{2} } \bigg) [\eta, \rho^0 ] \nonumber\\&&+
   {1 \over 3 \sqrt{2} } [K^+, K^{*+} ] -
   { \sqrt{2} \over 3}  [K^0, K^{*0} ] \nonumber \\
&&+ { \sqrt{2} \over 3} [\pi^+, \rho^+ ]  +
  \bigg( {s^2 \over 9 \sqrt{2} } -  {c s \over \sqrt{6} }
   \bigg)  [\pi^0, \omega ] +
   \bigg( {c^2 \over 3 \sqrt{2} }  -
   {c s \over \sqrt{6}} \bigg)  [\pi^0 , \rho^0 ] \bigg\} \nonumber \\
F_{\phi} &=& -\frac{\Lambda}{3}  \bigg\{
   {4 c^2 \over 3 } [\eta, \phi ] -
   [K^+, K^{*+} ] + 2 [K^0, K^{*0} ] +
   {4 s^2 \over 3} [\pi^0, \phi] \bigg\} \nonumber \\
F_{\rho^+} &=&  \Lambda
  \bigg\{
     {c^2 \over 3} [\eta, \rho^+ ]  + {1 \over 2 }[ K^+, K^{*0} ]
   + {1 \over 2 }[ K^0, K^{*+} ] +  [ \pi^+, \omega ]
   + { s^2 \over 3}  [\pi^0 , \rho^+ ]  \bigg\} \nonumber \\
F_{K^{*+}} &=& { \Lambda  \over 2 }
 \bigg\{
 {1\over 2} \bigg( {c \over \sqrt{3}} -s \bigg )^2
 [\eta, K^{*+} ]
+ {1 \over 2} [K^+, \omega]
+  [K^+, \phi]
+ {1 \over 2} [K^+, \rho_0] \nonumber  \\
&& + [K^0, \rho^+]
+  [\pi^+, K^{*0}]
+ {1 \over 2} \bigg(  c + {s \over \sqrt{3} } \bigg)^2
[ \pi^0, K^{*+}]
\bigg\}
\ea
with
\be
\Lambda = { 16 g \lambda_5 \over F_\pi^2 }, \quad
\cos \theta = c \quad and \quad
 \sin \theta = s.
\ee

In addition to Eq. (\ref{tree}), Eq. (\ref{tadpole}) and Eq. (\ref{sunri})
one
has the w.f.r. terms.

We have defined the following integrals
\ba
\label{int}
&&i(K g_{\mu \nu} + L v_\mu v_\nu) = \int\frac{d^dq}{(2\pi)^d}
\frac{ q_\mu q_\nu }
   {v\cdot q - \omega + i\eta}\frac{1}{q^2-m^2+i\eta}
\nonumber\\&&
i\mu_m=\int \frac{d^dq}{(2\pi)^d}\frac{1}{q^2 - m^2} =
\frac{i m^2}{16\pi^2}
\left[\lambda-\log\left(\frac{m^2}{\mu^2}\right)\right]\, ,
\ea
with $\lambda = 1/\epsilon-\gamma+\log(4\pi)+1$ and $d=4-2\epsilon$,
and
\be
\label{int2}
[G,\Omega] = \left( 1+{p\cdot v \over m_V} -{m_G^2 \over 2 m_V}
{\partial \over \partial m_\Omega } \right)
K(m_G,m_\Omega -m_V - {p\cdot v} ) \nonumber\\
\ee
Here $\eta$ and $\pi_0$ stand for the physical pseudoscalar fields.

\end{document}